\documentclass[twocolumn,prp,aps]{revtex4}
\usepackage{amsmath}
\usepackage{amssymb}
\usepackage{graphicx}
\usepackage{dcolumn}
\usepackage{graphicx}
\usepackage{dcolumn}
\usepackage{bm}

\begin{document}

 
\title{Microwave Photoresistance of a Two-Dimensional Topological Insulator in a HgTe Quantum Well}

\author{A. S.Yaroshevich,$^{1}$ Z. D. Kvon,$^{1,2}$ G. M. Gusev,$^3$ and N. N. Mikhailov,$^{1,2}$  }

\affiliation{$^1$Institute of Semiconductor Physics, Novosibirsk
630090, Russia}

\affiliation{$^2$Novosibirsk State University, Novosibirsk 630090,
Russia}

\affiliation{$^3$Instituto de F\'{\i}sica da Universidade de S\~ao
Paulo, 135960-170, S\~ao Paulo, SP, Brazil}

\date{\today}

\begin{abstract}
The microwave photoresistance of a two-dimensional topological insulator in a HgTe quantum well with an
	inverted spectrum has been experimentally studied under irradiation at frequencies of 110–169 GHz.
	Two mechanisms of formation of this photoresistance have been revealed. The first mechanism is due to transitions
	between the dispersion branches of edge current states, whereas the second mechanism is caused by the
	action of radiation on the bulk of the quantum well.

\end{abstract}

\maketitle

 A two-dimensional topological insulator in a HgTe quantum well with an inverted spectrum is the most
 reliable experimental implementation of a two-dimensional topological insulator; for this reason, it
 has been actively studied for more than a decade. A number of phenomena in it from transport and noise
 phenomena \cite{1Konig,2Roth,3Gusev,4Olshanetsky,5Tikhonov,6Kononov,7Piatrusha} to photoelectric
 phenomena such as a photogalvanic effect induced by the appearance of
 chiral spin photocurrents and a terahertz photoresistance caused by optical transitions between helical
 edge states \cite{8Dantscher,9Kvon} have already been studied.

In this work, we study for the first time the photoresistance of a two-dimensional topological insulator
appearing at microwave irradiation in the frequency range of 110-169\,GHz. The experimental samples
were microstructures with a special Hall geometry equipped with a semitransparent TiAu gate (see
Fig. 1a). They were fabricated on the basis of 8 to 8.5 nm thick HgTe quantum wells, where an inverse
energy spectrum is implemented. The structures were described in detail in \cite{9Kvon}.
The samples were irradiated by microwave radiation in the frequency range of 110-169\,GHz through a waveguide.
The radiation intensity at the output of the waveguide was in the range of 10-100\,mW/cm$^{2}$.
The photoresistance was measured by the double synchronous detection method at a radiation modulation
frequency of 22\,Hz at the passage of a current at a frequency of 1.1\,Hz and a magnitude 10--100\,nA through
the sample, which excludes the heating of electrons. We also note that the possible capacitance contribution
to a photoelectric signal and the corresponding distortion of its phase at resistances measured in the experiment
(below 1\,M$\Omega$) and used modulation frequencies (22\,Hz) were negligibly small because of a low capacitance
of the structure (about 0.1\,pF). The experiments were performed in the temperature range of 1.5-30\,K with
two groups of samples: with diffusion and quasiballistic transport. Both groups of the samples were fabricated
on the same initial quantum well and by the same technology. The mobility of electrons in the quantum well at their
density $n_s = 10^{11}$~cm$^{-2}$ was $\mu = 3 \times 10^4$~cm$^2$/(V\,s).
The samples were classified as diffusion and quasiballistic according to the measured local resistance
\emph{R}$^{L}_{14,65}$ (i.e., the resistance of the shortest bridge with the length \emph{L}\,=\,3.2\,$\mu$m, when
the current \emph{I} is passed between contacts 1 and 4 and the voltage \emph{V} is measured between
contacts 6 and 5) at the maximum of its dependence on the gate voltage \emph{V}$_{g}$.
The ballistic regime is implemented if , \emph{R}$^{max}_{14,65}$ = \emph{h}/2\emph{e}$^{2}$,
whereas the diffusion regime occurs when \emph{R}$^{max}_{14,65}$ $\gg$ \emph{h}/2\emph{e}$^{2}$.
It is noteworthy that the samples based on the same quantum well and fabricated by the same technology
demonstrate both the ballistic and diffusion regimes. This indicates that a universal topological protection
from backscattering is absent in real two-dimensional topological insulators.

\begin{figure}

\includegraphics[width=0.9\columnwidth]{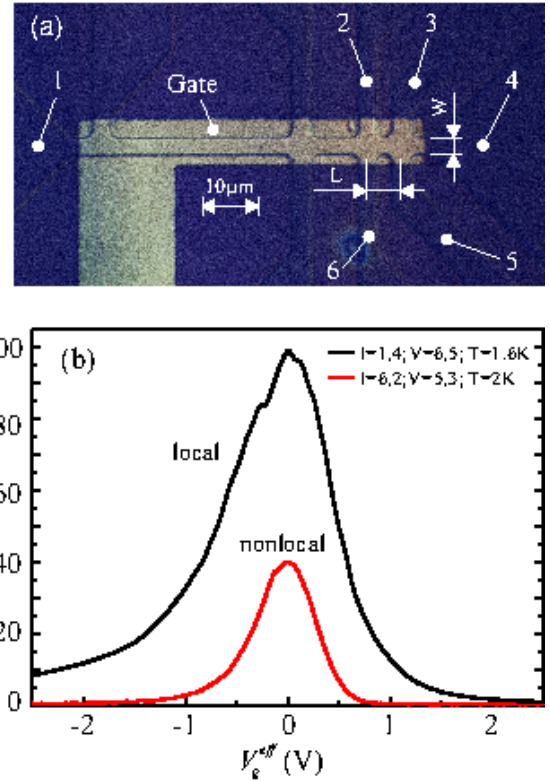}
\caption{
(Color\,online) (a) Photograph of the microstructure with a special Hall geometry.
(b)\,Local, \emph{R}$^{L}_{14,65}$(\emph{V}$^{eff}_{g}$),
and nonlocal, \emph{R}$^{nL}_{62,53}$(\emph{V}$^{eff}_{g}$), resistances versus the effective
gate voltage.
}
\label{fig1}
\end{figure}

We begin the description of the experiment with the results obtained for samples exhibiting the diffusion
transport regime, more precisely, with the analysis of their transport response.
The local, \emph{R}$^{L}_{14,65}$(\emph{V}$^{eff}_{g}$),
and nonlocal, \emph{R}$^{nL}_{62,53}$(\emph{V}$^{eff}_{g}$) (\emph{I}=6, 2; \emph{V}=5, 3),
resistance measured on the shortest part of the Hall bar
(see Fig. 1a) are shown in Fig. 1b as functions of the effective gate voltage \emph{V}$^{eff}_{g}$ =
\emph{V}$_{g}$ $-$ \emph{V}$_{g}^{max}$ (where \emph{V}$_{g}$ is the
gate voltage and \emph{V}$_{g}^{max}$ is the gate voltage corresponding to the maximum resistance).
The observed picture qualitatively coincides with that obtained for all two-dimensional topological
insulators in HgTe quantum wells exhibiting the regime mentioned above \cite{10Gusev}. The resistance is
low at gate voltages corresponding to the position of the Fermi level \emph{E}$_{F}$ in the conduction band
and passes through a maximum of 100 and 40 k$\Omega$ for the local and nonlocal resistances, respectively,
at the charge neutrality point, while the Fermi level passes through the middle of the gap. It is noteworthy
that, when the Fermi level appears in the allowed band, the nonlocal resistance becomes
negligibly low compared to the local resistance, as should be. Figure 2 shows the
same dependences at various temperatures. It is clearly seen that the resistance for both transport
configurations increases with decreasing temperature. The inset shows the temperature dependences
of the maximum local and nonlocal resistances. As seen, these dependences in the temperature
range of 10--30 K are exponential, which indicates the freezing of the bulk; the activation energies
for local and nonlocal geometries are 80 and 200 K, respectively. The difference between these
energies is most probably due to the inhomogeneity of the bulk gap in the plane of the quantum well,
which can be attributed to the inhomogeneity of both the thickness of the quantum well and the
degree of disorder. At \emph{T}\,<\,10\,K, an increase in the resistance becomes much slower
and is described by a power law. Such a behavior of the edge resistance was detected
and described in \cite{11Gusev} and is due to backscattering induced by the electron–electron
interaction \cite{12Kainaris}.

\begin{figure}

\includegraphics[width=0.95\columnwidth]{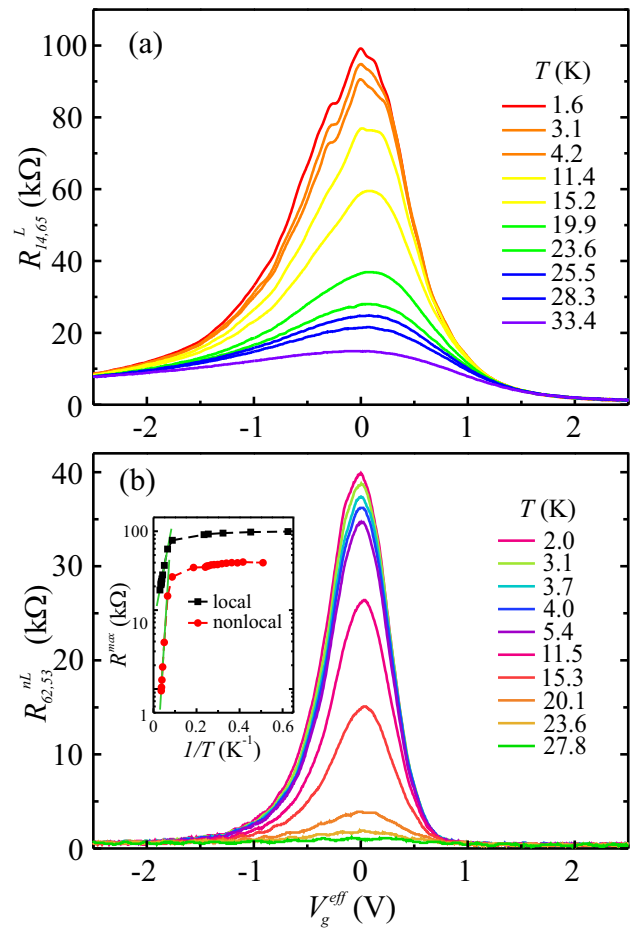}
\caption{
(Color\,online) (a) Local, \emph{R}$^{L}_{14,65}$(\emph{V}$^{eff}_{g}$), and (b)\,nonlocal, \emph{R}$^{nL}_{62,53}$(\emph{V}$^{eff}_{g}$), resistances versus the effective gate
voltage at various temperatures. The inset shows the
heights of the peaks \emph{R}$^{max}_{14,65}$ and \emph{R}$^{max}_{62,53}$ versus the inverse
temperature.
}
\label{fig2}
\end{figure}

Figure 3 shows the gate voltage dependences of the photoresistances \emph{$\vartriangle$R}$^{L}_{14,65}$(\emph{V}$^{eff}_{g}$) and
\emph{$\vartriangle$R}$^{nL}_{62,53}$(\emph{V}$^{eff}_{g}$) measured in the
local and nonlocal geometries, respectively. It is clearly seen that the photoresistance exists in both geometries. The nonlocal signal is comparable with the local one, which certainly indicates that the observed microwave response is due to change in the character  of the  edge transport. The sign of the observed photoresistance is negative; i.e., irradiation
reduces the resistance. In addition, Fig.\,3 clearly indicates that \emph{$\vartriangle$R}$^{L}$(\emph{V}$^{eff}_{g}$) and
\emph{$\vartriangle$R}$^{nL}$(\emph{V}$^{eff}_{g}$) reach maxima at
the same points as \emph{R}$^{L}$(\emph{V}$^{eff}_{g}$) and
\emph{R}$^{nL}$(\emph{V}$^{eff}_{g}$). Thus, the maximum photoresistance signal is observed when the Fermi level is located at the middle of the band gap, i.e., when transport is completely due to edge states. Furthermore, the comparison of the half-widths of the photoresistance and resistance curves shows that the gate voltage dependences of the photoresistance for the local response is twice as narrow as the same dependences for the resistance. All these properties indicate that the observed photoresistance is independent of  the effects of heating of the system, but is due to the photoexcited one-dimensional Dirac fermions appearing at transitions between edge branches.
The possibility of intense dipole transitions between these branches has been proven in recent theoretical work \cite{13Durnev}. To demonstrate a nonheating character of the detected photoresistance, Fig. 3 shows the gate voltage dependences of the photoresistance in comparison with the same dependences for the heating addition to the resistance determined from the temperature dependences of the local and nonlocal resistances of the studied topological insulator (Fig. 2). It is clearly seen that the half-width of the photoresistance curves in the local configuration is noticeably (by a factor of two) smaller than that for the heating addition, whereas the half-widths are the same for both dependences in the nonlocal geometry, when both dependences reflect just the edge transport. In fact, the heating addition repeats the gate voltage dependence of the resistance. The above analysis provides the certain conclusion that the found photoresistance of the two-dimensional topological insulator is determined by the excitation of nonequilibrium Dirac fermions at transitions between helical edge branches.

\begin{figure}

		\noindent \includegraphics[width=0.9\columnwidth]{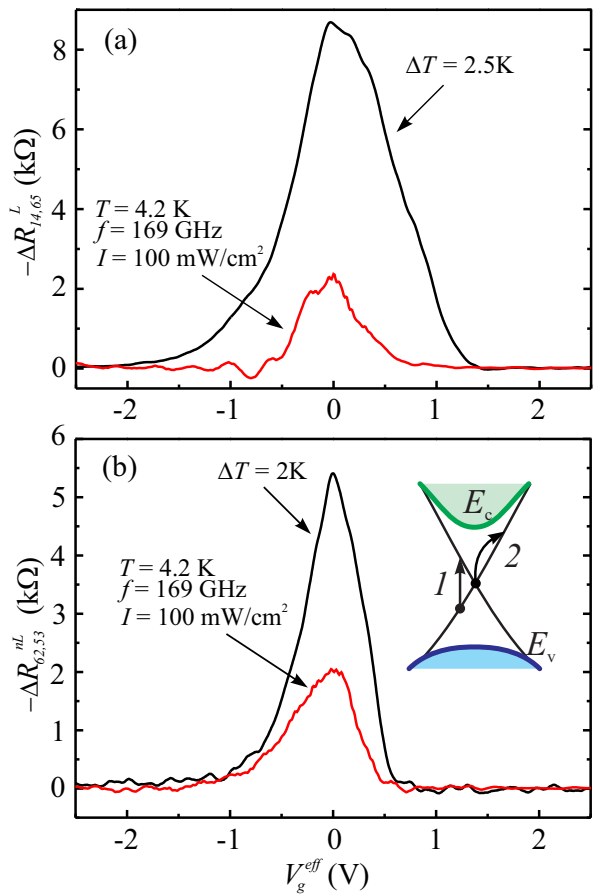}
	\caption
		{(Color online) (a) Local, \emph{$\vartriangle$R}$^{L}_{14,65}$(\emph{V}$^{eff}_{g}$), and (b)\,nonlocal,
		\emph{$\vartriangle$R}$^{nL}_{62,53}$(\emph{V}$^{eff}_{g}$), photoresistances versus the effective gate voltage in comparison with the heating addition to the resistance \emph{$\vartriangle$R}$^\emph{$\vartriangle$T}$(\emph{V}$^{eff}_{g}$).
		The inset illustrates (\emph{1}) direct
		transitions and (\emph{2}) transitions caused by Drude absorption.}
	\label{fig3}
\end{figure}

\begin{figure*}
	\includegraphics[width=1.9\columnwidth]{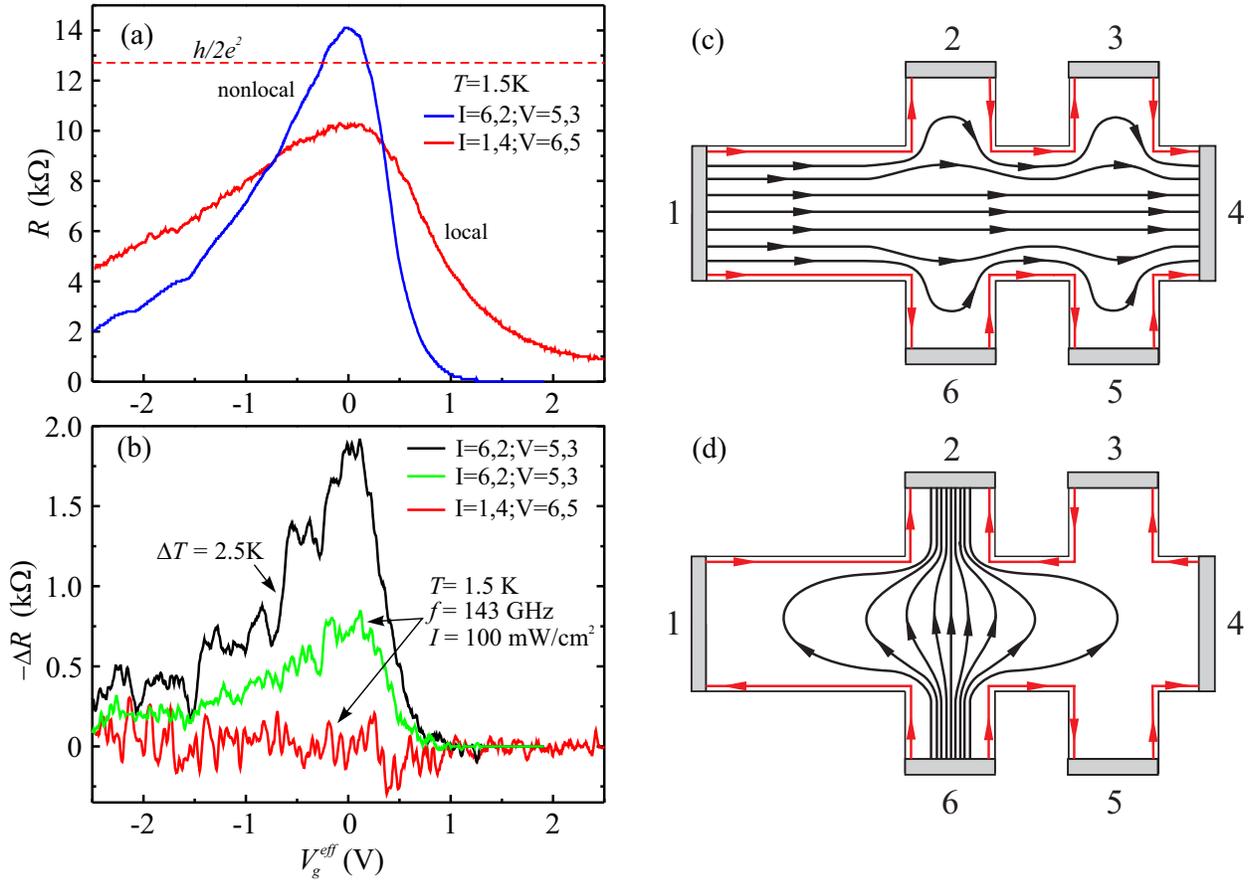}
	\caption{ (Color online) (a) Local, \emph{R}$^{L}_{14,65}$(\emph{V}$^{eff}_{g}$), and nonlocal, \emph{R}$^{nL}_{62,53}$(\emph{V}$^{eff}_{g}$), resistances versus the effective gate voltage.
		(b)\,Local, \emph{$\vartriangle$R}$^{L}_{14,65}$(\emph{V}$^{eff}_{g}$), and  nonlocal, \emph{$\vartriangle$R}$^{nL}_{62,53}$(\emph{V}$^{eff}_{g}$),
		photoresistances versus the effective gate voltage in comparison
		with the heating addition  to the nonlocal resistance \emph{$\vartriangle$R}$^\emph{$\vartriangle$T}_{62,53}$(\emph{V}$^{eff}_{g}$).
		(c, d)\,Current distributions in the sample with bulk leakage at measurements in local
		and nonlocal geometries, respectively. The current lines through the bulk are given in  black, whereas the current of edge states is shown in red. We note that the current in the absence of bulk leakage flows only through edge states.} \label{fig4}
\end{figure*}

We now discuss the results obtained for samples with quasiballistic transport. Their transport characteristics in the local and nonlocal configurations are shown in Fig. 4a. The maximum local resistance is slightly lower than \emph{h}/2\emph{e}$^{2}$, which indicates the ballistic transport regime but in the presence of a small bulk leakage. The measurements in the nonlocal geometry confirm this fact (Figs. 4c and 4d show the current distributions in the sample at the measurements in the local and nonlocal geometries, respectively). As seen, the maximum nonlocal resistance is only slightly higher than the corresponding local resistance, whereas the corresponding ratio should be \emph{R}$^{nL}_{max}$(\emph{V}$^{eff}_{g}$)/\emph{R}$^{L}_{max}$(\emph{V}$^{eff}_{g}$)= 4 if transport were due to only edge states. Figure 4b shows the microwave resistance also measured in local and nonlocal geometries. This figure obviously demonstrates that the picture observed for described samples differs from that
observed for the samples of the first group: the microwave resistance (also negative) is observed only in the nonlocal geometry. The response of the sample to irradiation in the local geometry is absent at the existing noise level. Such a behavior of the microwave resistance for samples with ballistic transport indicates a specific reaction of the topological insulator having a small bulk leakage. This behavior can be explained as
follows. Radiation incident on the sample primarily changes the bulk conductivity because of the heating effect. Consequently, the magnitude of bulk leakage changes, but this change is so small that its contribution at measurement in local geometry is not observed because it is combined with the resistance of the edge channel. The change in the nonlocal resistance under a change in the bulk conductivity should be much larger. In local geometry, \emph{$\vartriangle$R}$^{L}$$\approx$\emph{$\vartriangle$$\rho$}$_{xx}$(\emph{R}$_{edge}$/\emph{$\rho$}$_{xx}$)$^{2}$,
whereas in nonlocal geometry, \emph{$\vartriangle$R}$^{nL}$$\approx$\emph{$\vartriangle$$\rho$}$_{xx}$\,exp(-$\pi$L/W),
which is an order of magnitude higher than \emph{$\vartriangle$R}$^{L}$. Here, \emph{R}$_{edge}$ is the edge resistance, \emph{$\rho$}$_{xx}$ is the resistivity of the  bulk, \emph{$\vartriangle$$\rho$}$_{xx}$ is the radiation-induced change in the bulk resistance, \emph{W} is the width of the microbar, and \emph{L} is the distance between the potentiometric contacts (see Fig. 1a). Just this effect is observed in the samples with
ballistic transport. It is noteworthy that the contribution from transitions between the edge branches was not determined most probably because it is much weaker than the observed bulk contribution. To conclude, we discuss microscopic mechanisms of the hotoresistance of our topological insulator caused by transitions involving the edge branches. They can be
caused by two processes: (i) aforementioned absorption induced by transitions between edge branches \cite{13Durnev} and (ii) Drude absorption by these branches (see the inset of Fig. 3b) \cite{14Entin}. The maximum of absorption caused by the former obviously corresponds to the position of the Fermi level coinciding with the Dirac point, i.e., the charge neutrality point. This is just the experimental behavior of the photoresistance. The Drude absorption will be proportional to the density of states of one-dimensional Dirac electrons and holes \cite{14Entin}, which is constant and independent of the energy. As a result, when the Fermi level passes through the gap, the photoresistance should be independent of the position of the Fermi level and, thereby, of the gate voltage, which completely contradicts the experiment. Thus, the experiment certainly indicates that the experimentally observed microwave photoresistance caused by the interaction of radiation with Dirac branches is due to optical transitions between Dirac branches, which, as shown in \cite{13Durnev}, have a dipole character and are allowed because of spatial symmetry breaking at the HgTe/CdHgTe interface.

This work was supported by the Russian Science Foundation (project no. 16-12-10041-P) and by the S$\tilde{a}$o Paulo Research Foundation (FAPESP, Brazil).

\end{document}